# USING BIAS OPTIMIAZATION FOR REVERSIBLE DATA HIDING USING IMAGE INTERPOLATION

Andrew Rudder, Wayne Goodridge and Shareeda Mohammed

Department of Computing and Information Technology, The University of the West Indies, St Augustine, Trinidad and Tobago

andrew.rudder@gmail.com  wayne.goodridge@sta.uwi.edu
shareeda.mohammed@gmail.com

**ABSTRACT**

*In this paper, we propose a reversible data hiding method in the spatial domain for compressed grayscale images. The proposed method embeds secret bits into a compressed thumbnail of the original image by using a novel interpolation method and the Neighbour Mean Interpolation (NMI) technique as scaling up to the original image occurs. Experimental results presented in this paper show that the proposed method has significantly improved embedding capacities over the approach proposed by Jung and Yoo.*

# 1. INTRODUCTION

Secret information exchange is a vital requirement of persons and institutions in society. Channelling secret information over public networks has proven to be very insecure. There is a great need for protective methods for sending secret information.

Cryptography is traditionally the first method implemented for protection. Cryptography however, has a few drawbacks. Encrypted data can be easily identified during transmission hence attracting unwanted attention to the packaged information. With increased computing power the possibility of cracking the cryptographic technique increases. Information hiding is an alternative strategy that can be used to protect sensitive secret information. While cryptography protects the content of messages data hiding conceals the existence of secret information. In general, information hiding (also called data hiding or data embedding) includes digital watermarking and steganography [1].

Data hiding technology prevents information from being detected, stolen or damaged by unauthorized users during transmission. The word stegano-graphy is a Greek word meaning "covered writing" the art of hiding secret information in ways that prevent detection [2]. The traditional method of encryption [3] can still be applied to the message and then a stegonographic approach used to ensure undetected delivery.

Information can be hidden in many ways. Hiding information may involve straight message insertion whereby every bit of information in the cover is encoded or it may selectively embed messages in noisy areas that draw less attention. Messages may also be dispersed in a random fashion throughout the cover data.

Data hiding techniques can be carried out in three domains [4], namely, spatial domain [5], compressed domain [6] [7] [8], and frequency domain [9]. Each domain has its own advantages and disadvantages with regard to hiding capacity, execution time, storage space and etcetera. The fundamental requirements of information hiding systems are good visual image quality, large hiding capacity, robustness, and steganographic security (i.e., statistically undetectable) [4].

To design a new data hiding system which achieves all these factors (i.e. good visual quality, high hiding capacity, robustness, and steganographic security) is a technically challenging problem. Thus, there are different methods in designing data hiding systems in the literature. Some of these methods are as follows. The first method is to increase hiding capacity; commonly referred to as the embedding capacity or payload; while maintaining a good visual quality or at the cost of lower visual quality [10]. This approach is suitable to applications where high hiding capacity is desired. The second method proposes to devise a robust data hiding scheme [11]. This design serves robust watermarking systems. The third method aims at enhancing visual quality whilst maintaining the same hiding capacity or at the cost of lower hiding capacity [12]. The fourth method aims to develop a data hiding scheme with high embedding efficiency [5]. This approach can increase the steganographic security of a data hiding scheme because it is less detectable by statistical steganalysis [13].

In the stegonographic process the message, usually called the secret message, is embedded into a cover object such as a picture, sound file, video or other medium. The modified cover object called the stego object is then sent to the receiver. For all intents and purposes what is seen from an onlooker is a normal transfer of a particular object. The receiver, who has knowledge of the embedding method, then extracts the message from the cover object. In steganography, the message is of importance not the cover object. There are two general types of embedding strategies namely reversible [14] [15] [16] and non-reversible [17] [18]. In a reversible strategy, when the stego object is received, the message is extracted, but also the original object is reconstructed. In the non-reversible strategy the original object cannot be reconstructed.

Reversible data hiding methods enable the exact recovery of the original host signal upon extraction of the embedded information. Celik et al [14] classify reversible data hiding methods into two types additive spread spectrum techniques and embedded by modifying selected features of the host signal.

Image interpolation [19] addresses the problem of generating high-resolution images from its low resolution. Interpolation techniques stretch the size of the image by using known data to estimate values at unknown points.

Jung and Yoo [20] proposed the use of image interpolation in the sphere of data hiding. The Jung and Yoo method, neighbour mean interpolation (NMI) uses the adjacent pixels when interpolating and the data hiding method generates pixels to fill in the blanks and hide secret messages.

This paper begins with an overview of the Jung and Yoo method. The proposed method will then be presented followed by the experimental results and conclusion.

## 2. RELATED WORK

### 2.1 Jung and Yoo's Method

Jung and Yoo [3] introduced a data hiding method that used interpolation. Figure 1 shows the outline of their proposed method. An input image sized w x h is scaled down to quarter (¼) of its initial size. This scaled image is now used as the original image; in other words, the new input image is of size w/2 x h/2. Interpolation is then used to scale up the original image to produce a cover image of size w x h before embedding occurs. The secret message is then embedded into the cover image to produce a stego-image.

On the receiving side, the secret message can be extracted from the stego-image and the original cover reconstructed.

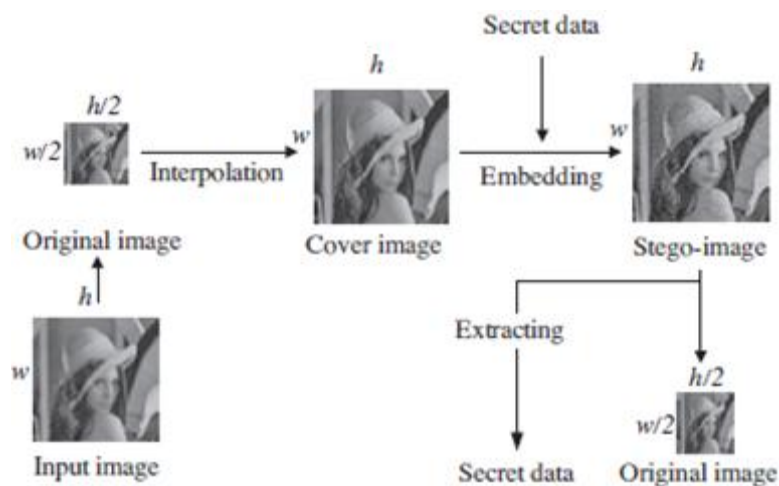

Figure 1: The flowchart of data hiding scheme by Jung and Yoo

#### 2.1.1 Neighbor mean interpolation

Jung and Yoo used the Neighbor Mean Interpolation method (NMI) to scale up the original image to produce the cover image *C* by embedding a secret message in *C*. NMI uses neighbouring pixel values in an image to obtain an average value that is then inserted into a pixel that has not been altered. The process of NMI uses is a s follows: Each 2 x 2 block in the original image is expanded into a 3 x 3 block in the cover image. The four corner pixels of the 3 x 3 image are equivalent to the four corner pixels of the 2 x 2 block. Values for each pixel in the 9 x 9 image are obtained using Equation (1) below.

$$C(i,j) = \begin{cases} I(i,j), & \text{if } i = 2m, j = 2n, \\ (I(i,j-1) + I(i,j+1))/2, & \text{if } i = 2m, j = 2n+1, \\ (I(i-1,j) + I(i+1,j))/2, & \text{if } i = 2m+1, j = 2n, \\ (I(i-1,j-1) + C(i-1,j) + C(i,j-1))/3, & \text{otherwise.} \end{cases} \quad (1)$$

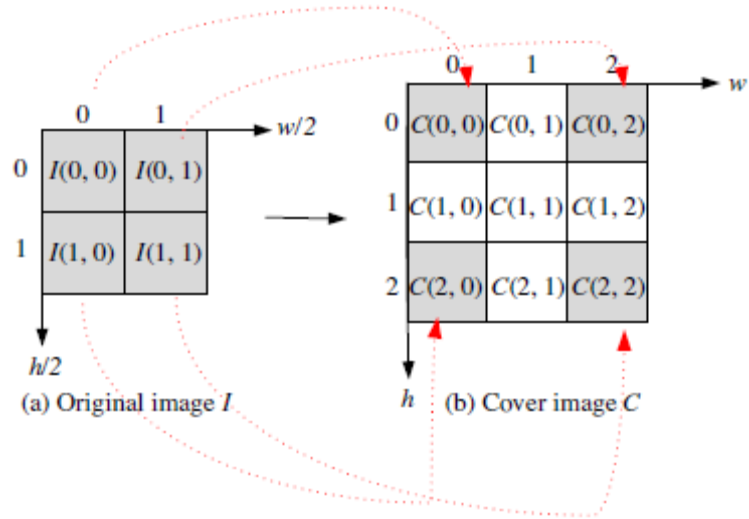

Figure 2: The cover image generated from the original image calculating NMI

A plus to this method is that it is applicable for interpoling computation when dealing with large images. This is due to the fact that good image quality and simple computation are two (2) advantages of NMI. The experimental results of Jung and Yoo demonstrated that the average image quality produced by NMI is 24.44db when measured by peak signal-to-noise-ratio.

### 2.1.2 Jung and Yoo's Embedding Scheme

In Jung and Yoo's scheme, a secret message, S, is embedded into a cover image C. The cover image is generated from the original image by using the NMI method. In order to achieve this, the cover image is partitioned into 2 x 2 non-overlapping blocks in a left-to right upper-to-lower zig-zag fashion. For each 2 x 2 block the pixel is in the top left corner C(i,j) carries the value from the original image and no embedding occurs. To pixels to the right C(i, j+1), below C(i+1, j) and diagonally down to the right C(i+1, j+1), all derived by the NMI method facilitate embedding. Secret data are embedded within three (3) pixels of a block except for the C(i,j) pixel.

To embed into a valid location C(p,q) where (p,q) $\in$ {(i, j+1), C(i+1, j) or C(i+1, j+1)}, the difference, d, is first calculated, where d is the C(p,q) – C(i,j). The number of bits that can be embedded into the cover pixel n is then calculated by the formula n = $\lfloor \log_2 d \rfloor$ where $\lfloor \rfloor$ represents the floor function. The next n bits of the secret message are read; let us call this sequence b. The stego pixel C'(p,q) is obtained from the cover pixel C(p,q) using the formula C'(p,q) = C(p,q) + b.

## 3. PROPOSED METHOD

In this section we propose a data hiding scheme which improves the embedding capacity of the cover image.

### 3.1 Interpolation procedure

The proposed method uses Neighbour Mean interpolation (NMI) in accordance with Jung and Yoo's scheme.

### 3.2 Embedding procedure

In our embedding scheme an interpolation method is proposed that embeds data as scaling up occurs. Once all data has been embedded, any remaining pixel values that need to be scaled up are done using Neighbour Mean Interpolation (NMI).

### 3.2.1 Embedding Method

The basic motivation for developing this technique is optimizing the embedding potential as well as generating pixels that have similar characteristics to their neighbouring pixels. Consider the original image I sized (w/2) × (h/2) and the binary secret message to be embedded *S*. The embedding method takes the interpolated cover image *C* sized w × h and produces a stego image *C'* sized w × h.

The cover image C is partitioned into 3 × 3 overlapping blocks of the cover image C by scanning in a zig-zag fashion from left to right and up to down directions.

The maximum and minimum values of selected neighbouring pixels from the original image I are calculated for each interpolated pixel that was generated.

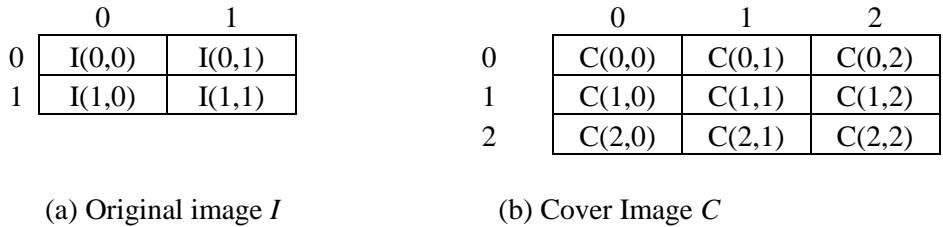

(a) Original image *I*      (b) Cover Image *C*

Figure 3: Cover image generated using NMI

The maximum value *Max* and minimum value *Min* for each interpolated pixel is calculated by Equation (2)

$Max(i, j) = \max\{I(i, j-1), I(i, j+1)\}$ if $i = 2m, j = 2n+1, 0 \leq j \leq i$ and $m, n = 0,1,2,…,127$.
$Max(i, j) = \max\{I(i-1, j), I(i+1, j)\}$ if $i = 2m + 1, j = 2n, 0 \leq j \leq i$ and $m, n = 0,1,2,…,127$.
$Max(i, j) = \max\{I(i-1, j-1), I(i-1, j+1), I(i+1, j-1)\}$ if $i = 2m + 1, j = 2n + 1, 0 \leq j \leq i$ and
$m, n = 0,1,2,…,127$.     (2)
$Min(i, j) = \min\{I(i, j-1), I(i, j+1)\}$ if $i = 2m, j = 2n+1, 0 \leq j \leq i$ and $m, n = 0,1,2,…,127$.
$Min(i, j) = \min\{I(i-1, j), I(i+1, j)\}$ if $i = 2m + 1, j = 2n, 0 \leq j \leq i$ and $m, n = 0,1,2,…,127$.
$Min(i, j) = \min\{I(i-1, j-1), I(i-1, j+1), I(i+1, j-1)\}$ if $i = 2m + 1, j = 2n + 1, 0 \leq j \leq i$ and
$m, n = 0,1,2,…,127$.     (3)

Once the *Max* and *Min* values have been calculated for an interpolated pixel, the difference value $d$ is calculated where $d = Max - Min$. We will denote $d_1$, $d_2$ and $d_3$ as the difference values for $C(i, j+1)$, $C(i+1, j)$ and $C(i+1, j+1)$ respectively.

The number of secret bits $n$ that can be embedded into an interpolated pixel is based on the value $d$ and is calculated as follows:

$$n_k = \lfloor \log_2 d_k \rfloor \text{ where } k = 1, 2, \text{ and } 3$$

The next $n$ bits of the secret message are read and its decimal equivalent *dec* is calculated and embedded using Equation 4.

$$C'(i, j) = C(i, j), \quad i = 2m, j = 2m$$
$$Min(i, j) + dec, \text{ otherwise}$$
$$\text{where } 0 \leq j \leq i \text{ and } m, n = 0, 1, 2, \ldots, 127. \quad (4)$$

Suppose a binary string for a secret message S, to be embedded in a cover image. Assume the pixel values for $I(0,0)$, $I(0,1)$, $I(1,0)$, $I(1,1)$ were 152, 161, 185 and 188 respectively. Let S = "$110011010111010100_2$". Using NMI the cover image $C$ seen in Fig 3 (b) would be generated.

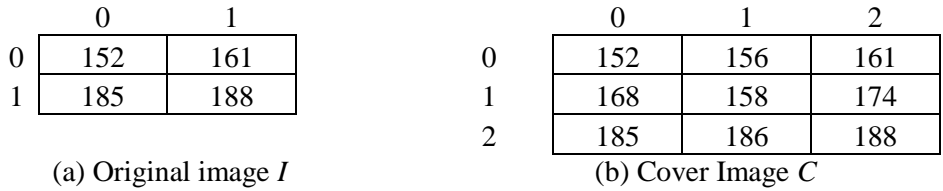

(a) Original image *I*      (b) Cover Image *C*

Figure 4: Cover image generated using NMI

The values for *Max* and *Min* are calculated for each interpolated pixel.

Max and min calculations for interpolated pixel $C(0,1)$

$Max_1 = \max\{C(0,0), C(0,2)\} = 161$

$Min_1 = \min\{C(0,0), C(0,2)\} = 152$

Max and min calculations for interpolated pixel $C(1,0)$

$Max_2 = \max\{C(0,0), C(2,0)\} = 185$

$Min2 = min\{C(0,0), C(2,0)\} = 152$

Max and min calculations for interpolated pixel $C(1,1)$

$Max_3 = \max\{C(0,0), C(2,0), C(0,2)\} = 185$

$Min3 = min\{C(0,0), C(2,0), C(0,2)\} = 152$

|     | C(0,1) | C(1,0) | C(1,1) |
| --- | --- | --- | --- |
| Max | 161 | 185 | 185 |
| Min | 152 | 152 | 152 |

Table 1: Summary of *Max* and *Min* values for $C(0,1)$, $C(1,0)$ and $C(1,1)$

The difference between their largest and smallest neighbouring pixels from the original image $I$ is then calculated as shown below.

$d_1 = Max_1 - Min_1 = 161 - 152 = 9$,

$d_2 = Max_2 - Min_2 = 185 - 152 = 33$, and

$d_3 = Max_3 - Min_3 = 185 - 152 = 33$.

The number of secret bits $n_1$, $n_2$ and $n_3$ that can be embedded into the interpolated pixels $C(0,1)$, $C(1,0)$ and $C(1,1)$ is calculated as follows.

$n_1 = \lfloor \log_2 d_1 \rfloor = 3$,

$n_2 = \lfloor \log_2 d_2 \rfloor = 5$, and

$n_3 = \lfloor \log_2 d_3 \rfloor = 5$

In the case where $n_1 = 3$, a 3 bit subset of the secret message is taken.

Let $bs_1$, $bs_2$ and $bs_3$ represent the binary bit sequences for $n_1$, $n_2$ and $n_3$ respectively. The secret bit string sequence $1100110101110101 00_2$ would now be broken up as follows.

$1100110101110101 00_2$ => $110\ \ 01101\ \ 01110\ \ 10100_2$

$\qquad\qquad\qquad\qquad\qquad\qquad bs_1 \quad bs_2 \quad\ \ bs_3$

Figure 5: Partitioning of the bit string

The decimal equivalents of $bs_1$, $bs_2$ and $bs_3$ are then added to the cover pixel values $C(0,1)$, $C(1,0)$ and $C(1,1)$. Let $d$ be a function that maps the binary values to decimal values. This will give.

$C'(0,1) = C(0,1) + d(bs_1) \quad = \ 152 + 6 \ = \ 158$

$C'(1,0) = C(1,0) + d(bs_2) \quad = \ 152 + 13 \ = \ 165$

$C'(1,1) = C(1,1) + d(bs_3) \quad = \ 152 + 14 \ = \ 166$

This is summarized in Figure 6 (b).

|   | 0 | 1 | 2 |
|---|---|---|---|
| 0 | 152 | 156 | 161 |
| 1 | 168 | 158 | 174 |
| 2 | 185 | 186 | 188 |

(a) Cover image $C$

|   | 0 | 1 | 2 |
|---|---|---|---|
| 0 | 152 | 158 | 161 |
| 1 | 165 | 166 | 171 |
| 2 | 185 | 185 | 188 |

(b) Stego image $C'$

Figure 6: Resulting stego pixels from the cover pixels with secret bit sequence $1100110101110101 00_2$

### 3.2.2 Extraction procedure

The extraction procedure involves partitioning the stego-image $C'$ into $3 \times 3$ overlapping blocks. For each stego pixel the *Max* and *Min* values are calculated using Equations 2 and 3 respectively.

To extract the secret bits embedded in a pixel, the number of bits embedded is firstly calculated. We achieve this by determining the difference value $d$ where $d = Max - Min$. We then calculate the length of the secret bit stream $n$, where $n = \lfloor \log_2 d \rfloor$ embedded into a stego pixel, . To extract the bits we subtract the stego pixel value from *Min* and then take the *n* bit binary equivalent.

A worked example on the stego image *C'* of Fig 6 (b) is demonstrated below.

Step 1:

The values for *Max* and *Min* are calculated for each embedded stego pixel.

Max and min calculations for stego pixel $C'(0,1)$

$Max_1 = \max\{C'(0,0), C'(0,2)\} = 161$

$Min_1 = \min\{C'(0,0), C'(0,2)\} = 152$

Max and min calculations for stego pixel $C'(1,0)$

$Max_2 = \max\{C'(0,0), C'(2,0)\} = 185$

$Min2 = \min\{C'(0,0), C'(2,0)\} = 152$

Max and min calculations for stego pixel $C(1,1)$

$Max_3 = \max\{C'(0,0), C'(2,0), C'(0,2)\} = 185$

$Min3 = \min\{C'(0,0), C'(2,0), C'(0,2)\} = 152$

Step 2:

The difference value *d* for the stego pixels $C'(0,1)$, $C'(1,0)$ and $C'(1,1)$ is calculated where:

$d_1 = Max_1 - Min_1 = 161 - 152 = 9$,

$d_2 = Max_2 - Min_2 = 185 - 152 = 33$, and

$d_3 = Max_3 - Min_3 = 185 - 152 = 33$.

Step 3:

The length of the secret bit streams $l_1$, $l_2$ and $l_3$ embedded into the stego pixels $C'(0,1)$, $C'(1,0)$ and $C'(1,1)$ is calculated as follows:

$l_1 = \lfloor \log_2 d_1 \rfloor = 3$,

$l_2 = \lfloor \log_2 d_2 \rfloor = 5$, and

$l_3 = \lfloor \log_2 d_3 \rfloor = 5$

Step 4:

The extracted decimal values $dec_1$, $dec_2$, $dec_3$ for each pixel $C'(0,1)$, $C'(1,0)$ and $C'(1,1)$ are calculated.

$dec_1 = 158 - 152 = 6$

$dec_2 = 165 - 152 = 13$

$dec_3 = 168 - 152 = 14$

Step 5:

The extracted decimal values are converted to their *l* bit binary equivalent and concatenated.

$dec_1$ is converted to its $l_1$ bit equivalent:     6 => $110_2$

$dec_2$ is converted to its $l_2$ bit equivalent:     13 => $01101_2$

$dec_3$ is converted to its $l_3$ bit equivalent:     14 => $01110_2$

The reconstructed bit steam:   110 || 01101 || 01110  => $1100110101110_2$

As can be seen, the extracted bit sequence $1100110101110_2$ is identical to the first 13 bits of the bits sequence embedded from Figure 6.

The original image $I$ can be reconstructed from the stego image $C'$ since the four corner pixels of $C'$ are the same as the pixels of the cover image $C$, which in turn is identical to the values of the $2 \times 2$ block in $I$. That is :

$C'(0,0) = C(0,0) = I(0,0)$

$C'(0,2) = C(0,2) = I(0,1)$

$C'(2,0) = C(2,0) = I(1,0)$

$C'(2,2) = C(2,2) = I(1,1)$

## 4. EXPERIMENTAL RESULTS

This section presents the The method proposed was implemented on an Intel Core i7-2600 CPU running at 3.4GHz with 4 GB RAM. Microsoft Visual Studio 2010 was used to develop the software in the C# language. The secret message $M$ of length $LM$ was randomly generated using the *random* class in C#.

The PSNR (peak signal to noise ratio) was used as an indicator of the quality of the stego image when compared to the original cover image. The bits per pixel (BPP) indicates the hiding capacity and is obtained by taking the total number of bits embedded in a given stego image and dividing it by the number of pixels in the stego image itself. Seven gray-scaled test images sized 512×512 were used as the cover images to test the proposed method. These images were evaluated in terms of image quality and embedding capacity.

Each 512x 512 grayscale image was scaled down to 256 x 256 and was treated as an original image which was scaled up once more to a size of 512 x 512 using NMI. The comparison results of image quality between each input image and the corresponding enlarged image are shown in Table 2.

The enlarged image was the cover image in which a secret message was embedded. The embedding message used the difference between the minimum and maximum value that is the range. The logarithm of this value correlates to the length of the secret message which can be embedded in the image. Due to the influence the neighbouring pixel the value tends towards the mean value.

Figure 12 shows the original, scaled and resultant images after the proposed method was applied. The proposed method produces images with less blurring and greater image resolution. The image quality of the proposed method is superior to that of Jung and Yoo's.

Further analysis of Table 2 indicates that the average BBP for the images on the Jung and Yoo's method is 0.78 as opposed to 1.54 for the proposed method. The proposed method almost doubles the embedding capacity while maintaining the image quality.

Figure 13shows clearly that the payload of the proposed scheme improved by an average of one hundred percent 100%) when the same PSNR values were used.

| Image Name | Original | Scaled Down | Resultant Image |
|---|---|---|---|
| a) Baboon | 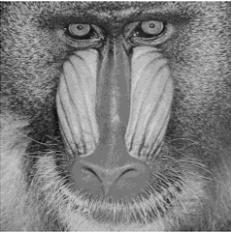 | 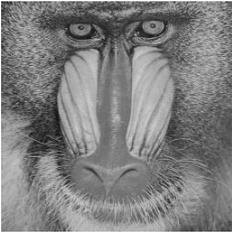 | 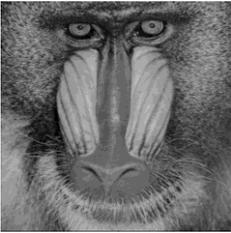 |
| b) Barbara | 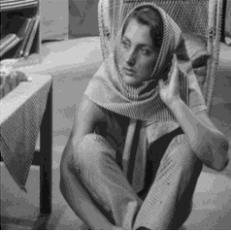 | 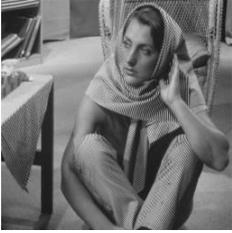 | 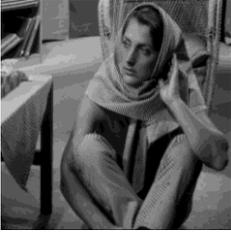 |
| c) Boat | 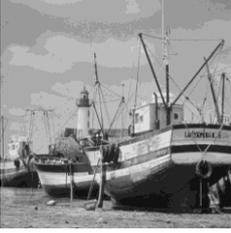 | 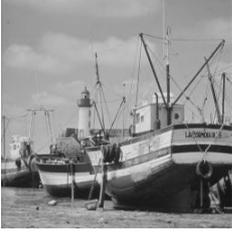 | 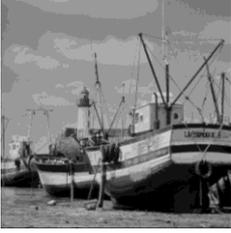 |
| d) F1 | 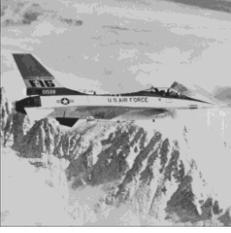 | 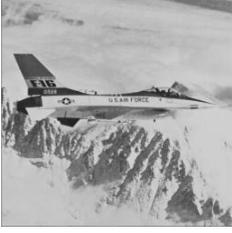 | 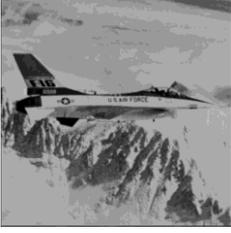 |
| e) GoldHill | 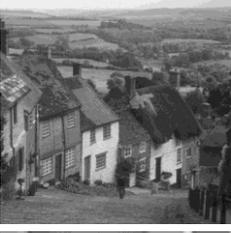 | 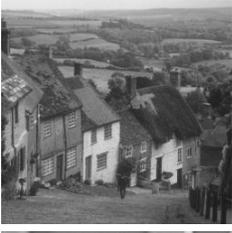 | 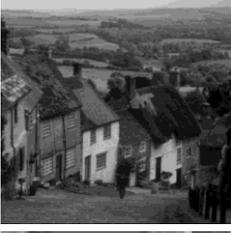 |
| f) Lena | 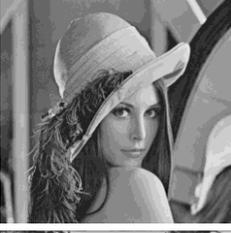 | 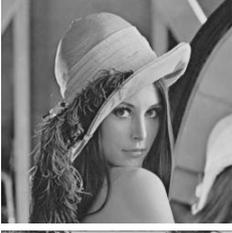 | 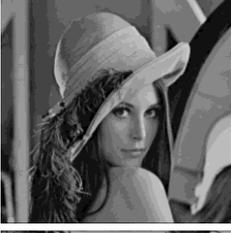 |
| g) Sailboat | 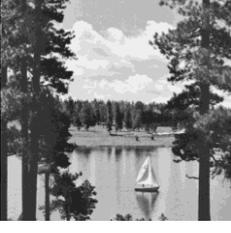 | 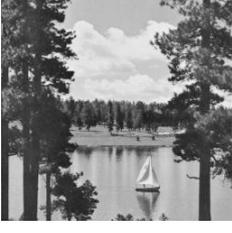 | 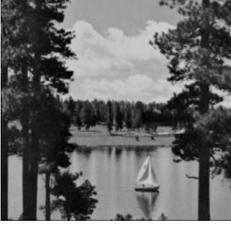 |

Figure 12: Grayscale cover images used in performance tests

| Image | Jung and Yoo | | | Proposed Method | |
|---|---|---|---|---|---|
| | PSNR | BBP | | PSNR | BBP |
| **Baboon** | 20.24 | 1.2634 | | 19.18 | 2.2286 |
| **Barabara** | 19.13 | 0.8989 | | 18.43 | 1.6923 |
| **Boat** | 20.77 | 0.605 | | 20.12 | 1.2214 |
| **Goldhill** | 19.09 | 0.7534 | | 18.55 | 1.5604 |
| **Jet** | 17.54 | 0.534 | | 17.14 | 1.133 |
| **Lena** | 19.95 | 0.5964 | | 19.42 | 1.2962 |
| **Sailboat** | 17.97 | 0.838 | | 17.33 | 1.6416 |

Table 2: Comparison of the proposed method with Jung and Yoo on the same cover images.

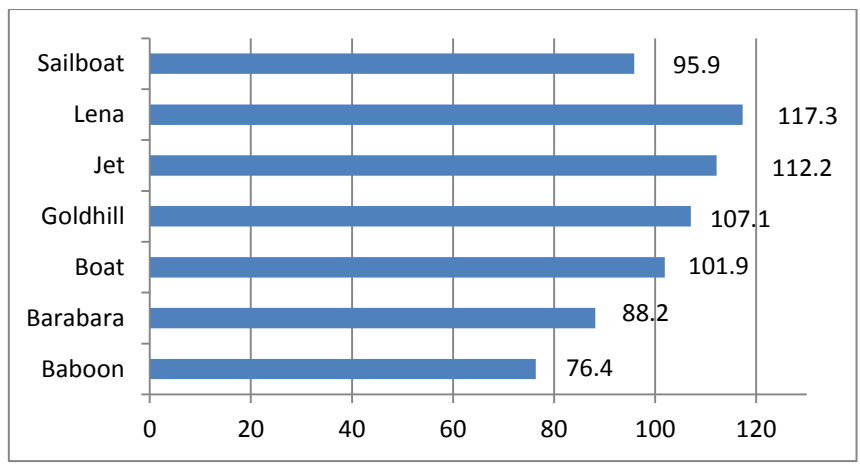

Figure 13: Gain rate of payload

## 5. CONCLUSION

This paper proposes a high-capacity image hiding scheme by exploiting NMI to improve the embedding capacity performance proposed by Jung and Yoo. It also offers the advantages of low computational complexity and good image quality.

The embedding method used neighboring pixels to establish the length of the secret message which can be embedded. The number of secret bits that can be embedded into an interpoled pixel is based on the difference of the maximum and minimum of the neighboring original pixels. The neighboring pixels generated had similar characteristics to each other. This influence allowed progression towards a median value in the number of pixels used. The result was that the embedding capacity was increased to almost double the capacity achieved by Jung and Yoo.

Reversible data hiding methods facilitate the recovery of the cover image after the secret data are extracted. Reversible data hiding methods are very complex and difficult to create. Our experimental results show the proposed method can embed a large amount of secret data while keeping a very high visual quality as demonstrated by the low PSNR values. In fact, with an embedding capacity that doubles the embedding capacity achieved by Jung and Yoo suggest that this algorithm could have very useful applications.